\documentclass[aps,prb,twocolumn,showpacs,letterpaper]{revtex4}  
\usepackage{mathrsfs}  
\usepackage{graphicx}  
\usepackage{dcolumn}   
\usepackage{bm}        
\usepackage{amssymb}   
\usepackage{afterpage} 

\usepackage{subfigure}
\usepackage{amsmath}

\begin{document}
\newcommand{\etal}{{\em et al.}}{}
\newcommand{\fig}[1]{Fig.~(\ref{#1})}
\newcommand{\tab}[1]{{Table ~(\ref{#1})}}
\newcommand{\eqn}[1]{{Eq.~(\ref{#1})}}
\newcommand{\angstrom}{\AA\,}{}    
\newcommand{\eVA}{$\rm{eV/\rm{\AA}}$\,}{} 
\newcommand{\ww}{0.8}



\title{Structure, mechanical and thermodynamic stability of vacancy clusters in Cu}

\author{Qing Peng, Xu Zhang and Gang Lu}

\affiliation{Department of Physics and Astronomy, California State
University Northridge, Northridge, CA, USA}

\date{\today}

\begin{abstract}
The atomic structure, mechanical and thermodynamic stability of
vacancy clusters in Cu are studied by atomistic simulations. The
most stable atomic configuration of small vacancy clusters is
determined. The mechanical stability of the vacancy clusters is
examined by applying uniaxial and volumetric tensile strain to the
system. The yield stress and yield strain of the system are
significantly reduced comparing to the prefect lattice. The
dependence of vacancy formation and binding energy as a function of
strain is explored and can be understood from the liquid-drop model.
We find that the formation energy of the vacancy clusters decreases
monotonically as a function of the uniaxial strain, while the
formation energy increases first then decreases under the volumetric
tensile strain. The thermodynamic stability of the vacancy clusters
is analyzed by calculating the Gibbs free binding energy and the
total probability of dissociation of the vacancy clusters at 300 K
and 900 K under uniaxial and volumetric strains. We find that
although most of the vacancy clusters appear to be thermodynamically
stable, some of the immediate sized clusters have high probability
of dissociation into smaller clusters.
\end{abstract}
\pacs{61.72.Bb,61.72.jd,61.72.J-,61.72.-y}
 \maketitle

\section{Introduction}
Vacancies are predominant point defects in metals and they could
have profound influence on materials properties, ranging from
mechanical strength \cite{Fisher1979,Zinkle1987,LuPRL2005},
deformation behavior \cite{Marian2004,LuPRL2002}, kinetic transport
and diffusion \cite{Seeger1970,Neumann1989}, to electrical
conductivity and heat capacity\cite{Kraftmakher1998,Kittel1996}, to
name a few. Thus the study of vacancy effects on properties of
technologically important materials, such as Cu, is of great
interest. For all these properties, stress plays crucial roles. For
example, it is generally believed that hydrostatic tensile stress is
the driving force for the stress-void nucleation \cite{Li2004} and
stress gradients are responsible for the void growth
\cite{Wang2006}. The present work touches upon the stress induced
voiding, which is one of the most important factors that cause
failures of ultra-large integrated circuits with Cu-based
interconnects \cite{Wu2008}. In this case, the stress is built up as
a result of the mismatch in thermal expansion coefficients between
the metal lines and the surrounding dielectrics \cite{Matsuea2006}.
There are other interesting problems related to the void growth and
coalescence under various loading conditions
\cite{Kolluri2008,Tszeng2008,Uberuaga2007,Rudd2002,
Ahn2006,Mukouda1999}. For example, experimentally TEM observations
have been performed for neutron-irradiated Cu at $300^\circ$C which
verify the transition from stacking fault tetrahedra (SFT) to voids,
depending on the size of the vacancy cluster. These observations are
consistent to the embedded-atom-method (EAM) simulations
\cite{Shimomura1993, Shimomura1997}. A direct transition from a
20-vacancy cluster to an SFT was also found by using parallel
replica EAM molecular dynamics simulations \cite{Uberuaga2007}. We
have recently studied void formation by using multiscale method and
discovered a connection between vacancy-induced bonding cage and the
tendency of void formation in fcc metals \cite{Zhang2008}.

In this paper, we will use atomistic simulations based on EAM
potential to (1) determine the structure of the most stable
$n$-vacancy clusters where $n\leq21$; (2) examine the mechanical
stability of Cu containing vacancy clusters; (3) study the
dependence of vacancy-cluster formation energy and binding energy as
a function of uniaxial and volumetric tensile stress; (4) examine
the thermal stability of the vacancy-clusters at different
temperatures. The computational method is given in Sec II. The
convergence of vacancy formation energy is studied in Sec III(A);
the structure of vacancy clusters is determined in Sec III(B) and
the mechanical stability of the vacancy clusters is examined in Sec.
III(C) for both uniaxial and volumetric tensile strains. Finally the
thermal stability of the vacancy clusters is considered in III(D)
and the conclusion is in Sec IV.

\section{Computational Method}
A cubic supercell with periodic boundary conditions along three
directions is used in the simulations. The supercell is oriented in
cubic directions, i.e., $x$ in $[100]$, $y$ in $[010]$ and $z$ in
$[001]$, respectively. The atomic interaction is described by
Ercolessi-Adams EAM potential\cite{eam}, which has been widely used
for the metallic systems, such as Cu
\cite{Kolluri2008,Tszeng2008,Uberuaga2007,Rudd2002,Ahn2006,Marian2004,Shimomura1993,Shimomura1997,Shimomura2003}.
The atomic relaxation is carried out with molecular statics
implemented in the LAMMPS codes\cite{Plimpton1995}. It is well
established that the EAM potential is capable to provide reasonable
atomic structure and energetics for vacancies in
Cu\cite{Kolluri2008,Uberuaga2007,Rudd2002}.

The vacancy cluster formation energy, which describes the energy
cost for forming a vacancy cluster with respect to the perfect
lattice, is computed for various vacancy clusters, in the presence
or absence of external strains. Specifically, under a strain
$\epsilon$ the vacancy cluster formation energy,
$E^{\rm{F}}_{n\rm{v},\epsilon}$ of a given $n$-vacancy cluster in an
$N$-atom supercell is defined as
\begin{equation}
\label{eq:Eforminge}
E^{\rm{F}}_{n\rm{v},\epsilon}=E^{\rm{tot}}_{n\rm{v},\epsilon}(N-n)-\frac{N-n}{N}E^{\rm{tot}}_{0,\epsilon}(N),
\end{equation}
where $E^{\rm{tot}}_{n\rm{v},\epsilon}(N-n)$ is the total energy of
the supercell containing ($N-n$) atoms with an $n$-vacancy cluster
under strain $\epsilon$ and $E^{\rm{tot}}_{0,\epsilon}(N)$ is the
total energy of the supercell containing $N$ atoms in a perfect
lattice with a strain of $\epsilon$. It is convenient to use the
formation energy per vacancy $E^{\rm{f}}_{n\rm{v},\epsilon}$ to
compare among different vacancy clusters, which is defined as
\begin{equation}
\label{eq:Eforminge2}
E^{\rm{f}}_{n\rm{v},\epsilon}=E^{\rm{F}}_{n\rm{v},\epsilon}/n.
\end{equation}
Another important quantity calculated is the vacancy binding energy,
$E^{\rm{B}}_{n\rm{v},\epsilon}$, which describes the energy gain
when $n$ monovacancies are combined into an $n$-vacancy cluster. The
binding energy of an $n$-vacancy cluster under a strain $\epsilon$
is given by
\begin{equation}
\label{eq:Ebindinge}
E^{\rm{B}}_{n\rm{v},\epsilon}=nE^{\rm{f}}_{1\rm{v},\epsilon}-E^{\rm{F}}_{n\rm{v},\epsilon}.
\end{equation}
A positive value of $E^{\rm{B}}_{n\rm{v},\epsilon}$ suggests that it
is energetically favorable for the $n$ monovacancies to form an
$n$-vacancy cluster at zero temperature under the strain. Similarly,
one can define the binding energy per vacancy,
$E^{\rm{b}}_{n\rm{v},\epsilon}$, to facilitate the comparison among
different vacancy clusters, which is given by
\begin{equation}
\label{eq:Ebinding2e}
E^{\rm{b}}_{n\rm{v},\epsilon}=E^{\rm{B}}_{n\rm{v},\epsilon}/n=E^{\rm{f}}_{1\rm{v},\epsilon}-E^{\rm{f}}_{n\rm{v},\epsilon}.
\end{equation}

In order to determine the thermodynamic stability of a vacancy
cluster, we need to calculate dynamical matrix and free energy of
the defect system. Under the harmonic approximation, the dynamical
matrix under strain $\epsilon$ is defined as following
\begin{equation}
D_{i,j}^{\alpha,\beta}=\frac{1}{m}\frac{\partial^{2}E_{\epsilon}^{\rm
tot}}{\partial{u}_{i}^{\alpha}\partial{u}_{j}^{\beta}},
\end{equation}
where $m$ is the mass of the atoms, $u_{i}^{\alpha}$ and
$u_{j}^{\beta}$ are the displacement of atom $i$ and $j$ in
direction $\alpha$ and $\beta$, respectively from its relaxed
equilibrium position. $E_{\epsilon}^{\rm tot}$ is the total energy
of the system. The vacancy formation entropy is evaluated via the
equation \cite{Carling2003,Mishin2001}
\begin{equation}
S_{n\rm{v},\epsilon}^{\rm F}= -k_{\rm B}
\left(\sum_{i=1}^{3(N-n-1)}{\rm ln}\omega_{i}^{n\rm
v,\epsilon}-\frac{N-n-1}{N-1}\sum_{i=1}^{3(N-1)}{\rm
ln}\omega_{i}^{0,\epsilon}\right),
\end{equation}
where $\omega_{i}^{n\rm{v},\epsilon}$ and $\omega_{i}^{0,\epsilon}$
represents the harmonic vibrational frequency for the system with
and without vacancies, respectively. Three acoustic phonon
frequencies are removed from the sum owing to the translational
invariance of the system. Analogous to the definition of the binding
energy, the binding entropy is defined as
\begin{equation}
S_{n\rm{v},\epsilon}^{\rm B}={n}S_{\rm 1v,\epsilon}^{\rm
F}-S_{n\rm{v},\epsilon}^{\rm F},
\end{equation}
where $S_{\rm 1v,\epsilon}^{\rm F}$ and $S_{n\rm{v},\epsilon}^{\rm
F}$ are the formation entropy of a monovacancy and an $n$-vacancy
cluster, respectively.

The dynamical matrix is determined by displacing the atoms
${\rm\pm0.01 \AA}$ from their relaxed equilibrium positions in all
three Cartesian directions. \cite{Carling2003} The dynamical matrix
is then diagonalized and the eigen-frequencies are obtained to
evaluate the formation entropy. The Gibbs binding free energy of an
$n$-vacancy cluster at temperature $T$ is determined as
\begin{equation}
G_{n\rm{v},\epsilon}^{\rm B}=E_{n\rm{v},\epsilon}^{\rm
B}-T{S}_{n\rm{v},\epsilon}^{\rm B},
\end{equation}
The positive (negative) value of $G_{n\rm{v},\epsilon}^{\rm B}$
indicates the $n$-vacancy cluster is thermodynamically stable
(unstable) against a simultaneous dissociation into $n$
monovacancies.

\section{Results and Analysis}
\subsection{Convergence of vacancy formation energy}
\begin{figure}[htp]
\includegraphics[width=\columnwidth]{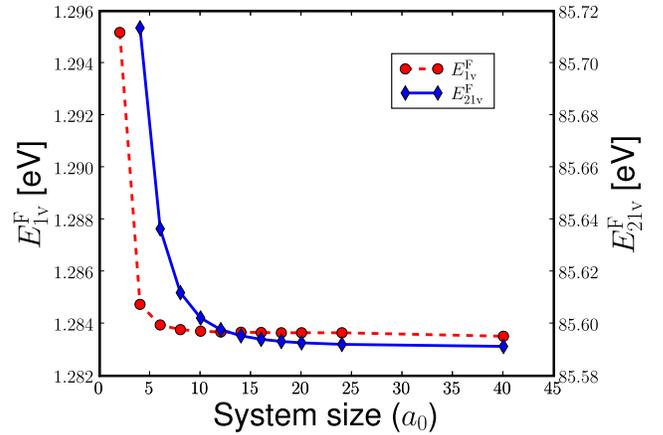}
\caption{(Color online) The convergence test of the vacancy
formation energy with respect to the system size for a monovacay
(dashed line, scaled to the left axis) and a 21-vacancy
cluster(solid line, scaled to the right axis).}\label{fig:E1f}
\end{figure}

A convergence test is carried out to determine the sufficient system
size, and results are shown in \fig{fig:E1f}. For a monovacancy, we
find 6$a_0$ ($a_0=3.615$ \AA) is enough to reach convergency for the
vacancy formation energy, while for $21$-vacancy cluster (the
largest vacancy cluster in this study), 20$a_0$ appears to be
sufficient. Therefore, in all our calculations, we use 20$a_0$ as
the system size. With a supercell of $20a_0 \times 20a_0 \times
20a_0$ (32000 lattice sites), we found that the monovacancy
formation energy is 1.2837 eV, which agrees very well with previous
calculation and experiment\cite{Nordlund1997,Anderson2004}.

\subsection{Atomic structure of $n$-vacancy clusters}

The vacancy clusters containing up to twenty-one vacancies are
studied in this work. For each $n$-vacancy cluster, several most
stable configurations are considered, especially in planar,
tetrahedral, and spherical structures. These structures are
emphasized owing to their relevance to vacancy loops, stacking fault
tetrahedra and spherical voids, observed in
experiments\cite{Zinkle1987}. For each $n$, the structure with the
maximum binding energy is regarded as the most energetically stable
configuration for the $n$-vacancy cluster.

The most stable di-vacancy is the one that the two vacancies are the
nearest neighbors. The binding energy is 0.158 eV, agreeing well
with the previous calculation and
experiment\cite{Nordlund1997,Anderson2004}. Starting from the most
stable di-vacancy, we can search for the most stable tri-vacancy by
placing the third vacancy at various nearest-neighbor position. The
configuration of the highest binding energy of a tri-vacancy is
shown in Fig. 2(a) and $E^{\rm{B}}_{3v}=0.474$ eV. The three
vacancies form an equilateral triangle lying on \{111\} plane, and
are mutually nearest-neighbors. Starting from the most stable
tri-vacancy, we considered six most probable configurations for a
four-vacancy cluster, and found that the equilateral tetrahedron
formed by four nearest-neighbor vacancies has the highest binding
energy of  $E^{\rm{B}}_{4v}=0.94$ eV. The atomic structure of the
tetra-vacancy is shown in Fig. 2(b).

\begin{figure}
\centering
\mbox{\subfigure{\includegraphics[width=\columnwidth]{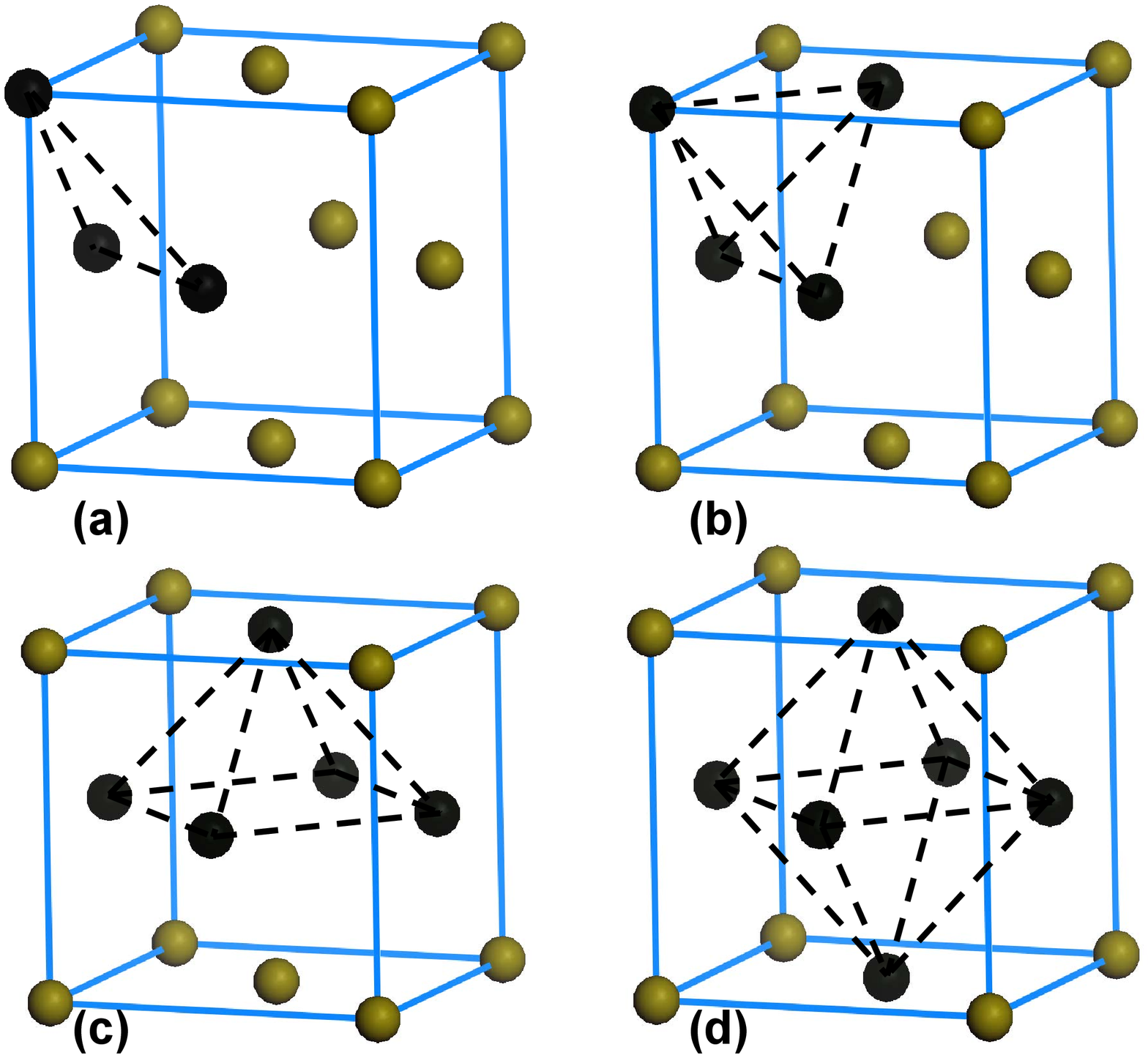}}}
\mbox{\subfigure{\includegraphics[width=\columnwidth]{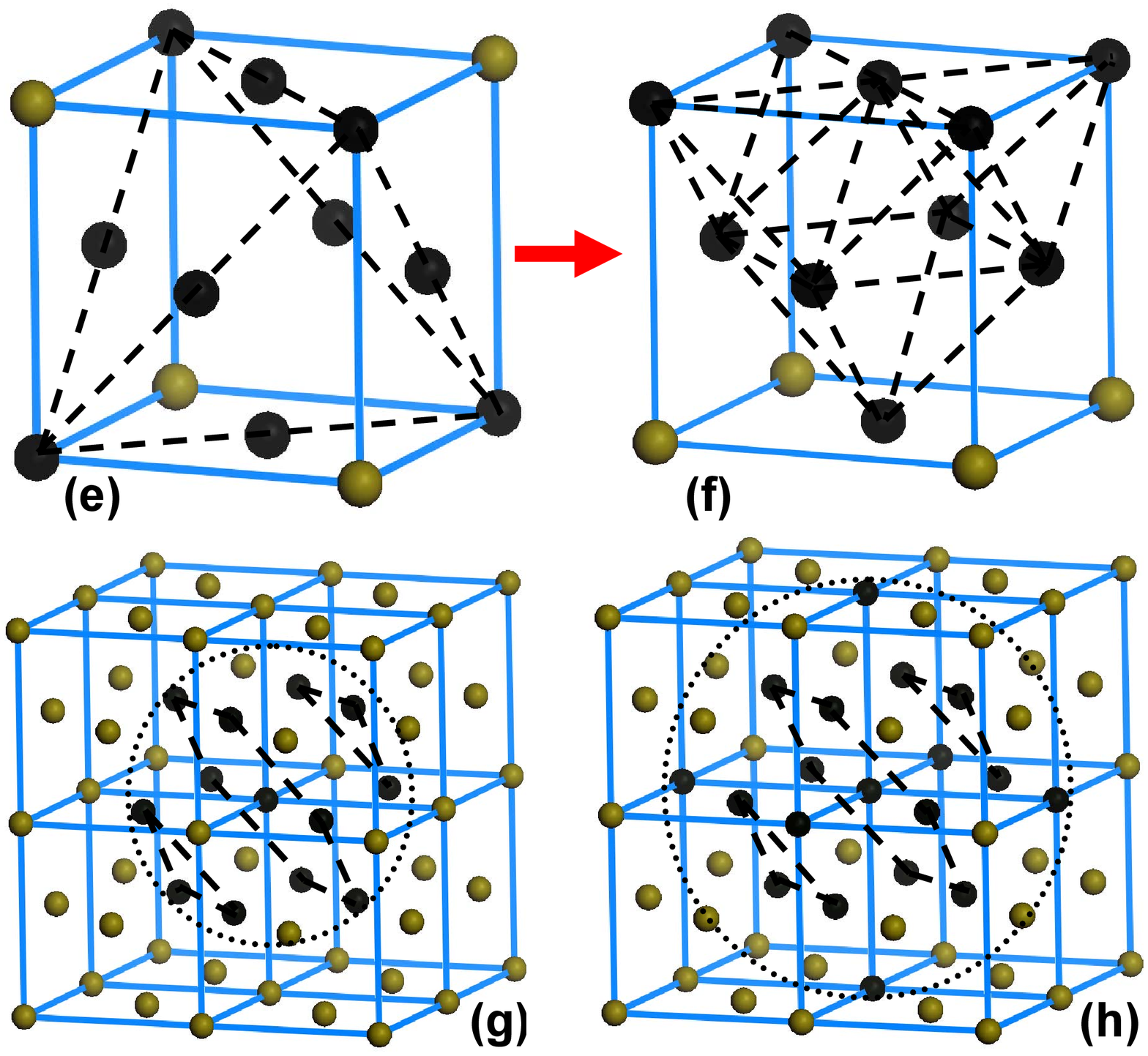}}}
\caption{The most stable atomic configurations for 3, 4, 5, 6, 10,
13, and 19-vacancy clusters. The black circle represents for the
vacancy and the brown circle for the atoms. Note that the most
stable configuration for the 10-vacancy cluster is not the
symmetrical one shown in (e), but rather the one with less symmetry
shown in (f).} \label{fig:v3}
\end{figure}

In the same manner, we explored the most stable atomic
configurations for all vacancy clusters with $n \leq 21$. For
example, we found the most stable 5-vacancy cluster (penta-vacancy)
is a squared pyramid with a binding energy of
$E^{\rm{B}}_{5v}=1.372$ eV, shown in Fig. 2(c). The most stable
6-vacancy cluster (hexa-vacancy) is an octahedron shown in Fig. 2(d)
with a binding energy of $E^{\rm{B}}_{6v}=2.113$ eV. Interestingly,
we found that the most stable configuration of a 10-vacancy cluster
is not an equilateral tetrahedron shown in Fig. 2(e), but rather a
less symmetric structure shown in Fig. 2(f). For $n=13$, the most
stable structure resembles a sphere, with one vacancy at the center,
and 12 nearest neighbors surrounding it. The radius of the sphere is
$\sqrt{2}/2 a_0$ shown in Fig. 2(g). For $n>13$, the vacancy
clusters tend to form 3-dimensional voids, appearing more like
spheres than tetrahedra. For example, the structure of a 19-vacancy
cluster is shown in Fig. 2(h); it has one vacancy at center, and 12
nearest neighbors and 6 second-nearest neighbors of the central
vacancy filling the sphere of a radius of $a_0$.

\begin{figure}[htp]
\includegraphics[width=\columnwidth,angle=0]{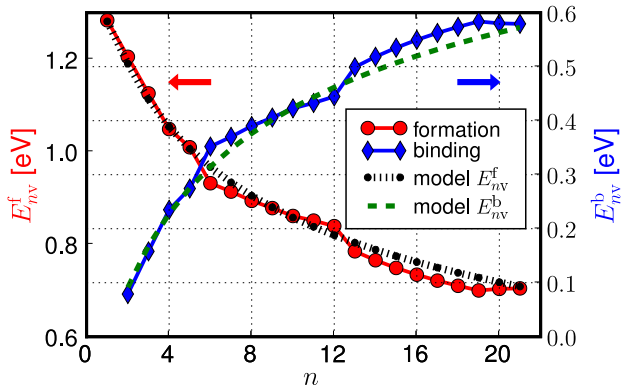}
\caption{The formation energy (circle, scaled to left) and binding
energy (diamond, scaled to right) per vacancy of the $n$-vacancy
clusters. The dot-dashed line and dashed line are the prediction of
$E^{\rm{f}}_{n\rm{v}}$ and $E^{\rm{b}}_{n\rm{v}}$ from the
liquid-drop model respectively.}\label{fig:bindinga}
\end{figure}

The formation energy per vacancy $E^{f}_{n\rm{V}}$ for all vacancy
clusters is presented in Fig. 3.  Overall, the formation energy
decreases with respect to the cluster size, which can be understood
from the liquid-drop model for metals \cite{Perdew91,Carling00}. In
this model, the void formation energy can be approximated as
\begin{equation}
\label{eq:liquid} E^{\rm{F}}_{\rm{void}}=4 \pi R^2_{\rm WS} \sigma -
2 \pi R_{\rm WS} \gamma,
\end{equation}
where $R_{\rm WS}=(3V/4 \pi )^{1/3}$ is the Wigner-Seitz radius of
the void with volume $V$, $\sigma$ is the surface energy density,
and $\gamma$ is the curvature energy \cite{Perdew91}. The volume of
an $n$-vacancy cluster can be approximated as $V_n \approx nv_0$,
where $v_0$ is the volume of a monovacancy. Therefor the
\eqn{eq:liquid} can be rewritten as
\begin{equation}
\label{eq:liqidn} E^{\rm{f}}_{n\rm{v}}=c_1 n^{-\frac{1}{3}} - c_2
n^{-\frac{2}{3}},
\end{equation}
where $c_1=4 \pi \sigma (3v_0/4 \pi)^{2/3}$, $c_2=2 \pi \gamma
(3v_0/4 \pi)^{1/3}$. Taking $\sigma =0.0932$ eV/\rm{\AA}$^2$,
$\gamma$ = 0.119 \eVA, and $v_0=a_0^3/4$, the Eq. (10) is plotted in
\fig{fig:bindinga}, which matches very well the atomistic results.
 The value of $\sigma$ and $\gamma$ is taken from
reference \cite{Perdew91}. $\sigma$ is close to the EAM surface
energy density of Cu: 0.08 eV/\rm{\AA}$^2$ for (110) surface and
0.114 eV/\rm{\AA}$^2$ for (111) surface. Interestingly, we find that
the formation energy curve becomes flatter as $n$ becomes larger,
and approaches an asymptotic value of ~ 0.7 eV. This value is very
close to the monovacancy formation energy on a flat (111) surface of
Cu, which is 0.73 eV. The reason for this asymptotic behavior is
that as the vacancy cluster becomes larger, the curvature of the
void surface becomes smaller, and thus the formation energy
approaches the value of a monovacancy on a flat surface.

We also present the binding energy per vacancy as a function of $n$
in Fig. 3, shown in diamond curve and scaled to the right axis. The
binding energy increases monotonically with $n$, and is positive for
all vacancy clusters, suggesting that the monovacancies prefer to
aggregate in Cu, in order to reduce the number of broken bonds. From
\eqn{eq:liqidn} and \eqn{eq:Ebinding2e}, we can express
$E^{\rm{b}}_{n\rm{v}}$ as
\begin{equation}
\label{eq:liqidb} E^{\rm{b}}_{n\rm{v}}= c_1(1-n^{-\frac{1}{3}}) -
c_2 (1-n^{-\frac{2}{3}}),
\end{equation}
which is plotted as the dashed line in \fig{fig:bindinga}. It
appears that the liquid-drop model reproduces the atomistic results
surprisingly well.

\subsection{Mechanical stability}
In this section, we examine the mechanical stability of the vacancy
clusters by applying external uniaxial tensile strain and volumetric
tensile strain to the system. The strain-free structures of the
vacancy clusters are taken from the results obtained in Sec. B.

\subsubsection{Uniaxial tensile strain}
The uniaxial tensile strain is applied along the $y$ axis, and the
$x$ and $z$ axes are free to change by conserving the volume. The
strain is applied quasi-statically, with each loading that is 1.001
times that of the previous step. We examine the mechanical response
of the vacancy clusters by showing the strain energy and tensile
stress $\sigma_{yy}$ as a function of applied strain
$\epsilon_{yy}$. Because the results of all vacancy clusters follow
the same pattern, we select the $13$-vacancy cluster as a
representative to illustrate the salient features of the results. As
shown in Fig. 4, we find that both strain energy and the tensile
stress increase monotonically with strain initially, followed by a
sudden drop at the yield point. The system behaves in a linear
elastic fashion, i.e., the energy increases quadratically and the
stress increases linearly. There are two importance results: (1) the
perfect and the defect lattice have almost identical elastic
behavior - the two curves coincide with each other for both energy
and stress. (2) The presence of the vacancy defect significantly
reduces the yield strength of the material. The yield stress and
strain are 7.32 GPa and 0.117 for the prefect lattice, and 4.98 GPa
and 0.085 for the defect system, respectively. At the yield point
(the drop in the curves), dislocations are nucleated and propagate
through the system, evidenced by the Central Symmetry Parameter
(CSP) \cite{Kelchner1998,Marian2004} plot in Fig. 5. CSP is defined
as $CSP=\sum\limits_{i=1}^{6}\vert R_i+R_{i+6} \vert^2/24D^2$, where
$R_i$ and $R_{i+6}$ are the vectors or bonds corresponding to the
six pairs of opposite nearest neighbors in the deformed fcc lattice
\cite{Kelchner1998}, and $D$ is the nearest neighbor distance in the
perfect fcc lattice. For each atom, there is a CSP value associated
with it. Dislocation core and stacking fault correspond to the CSP
values ranging between 0.003 and 0.1. Just before the yielding,
there is no dislocation in the system (Fig. 5a) while just after the
yielding, dislocations appear (Fig. 5b).
\begin{figure}[htp]
\includegraphics[width=\columnwidth,angle=0]{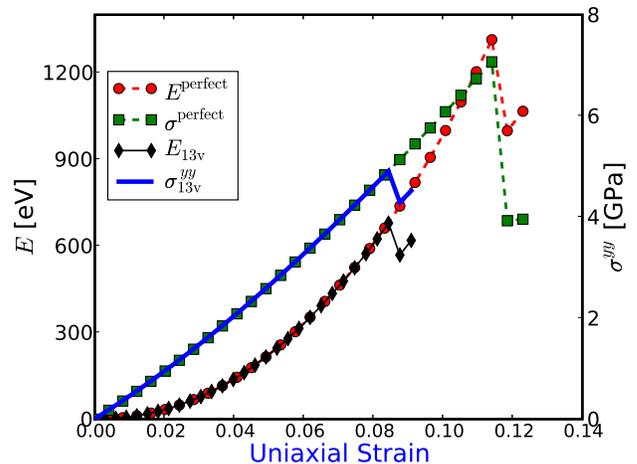}
\caption{The strain energy (scale to the left axis) and stress
(scale to the right axis) vs. uniaxial tensile strain for a
$13$-vacancy cluster (diamond and solid line) and perfect
lattice(circle and square) respectively.}\label{fig:v13strain}
\end{figure}

In Fig. 6, we present the yield strain (left axis) and yield stress
(right axis) as a function of $n$. Overall, the yield strain traces
approximately the yield stress, and for large vacancy clusters, the
yield stress/strain is significantly reduced comparing to the
perfect lattice. However, the vacancy effect on the mechanical
strength is more complex than the simple monotonic behavior. More
atomistic studies are required to unravel the complicated
vacancy-dislocation interactions.

\begin{figure}[htp]
\includegraphics[width=\columnwidth,angle=0]{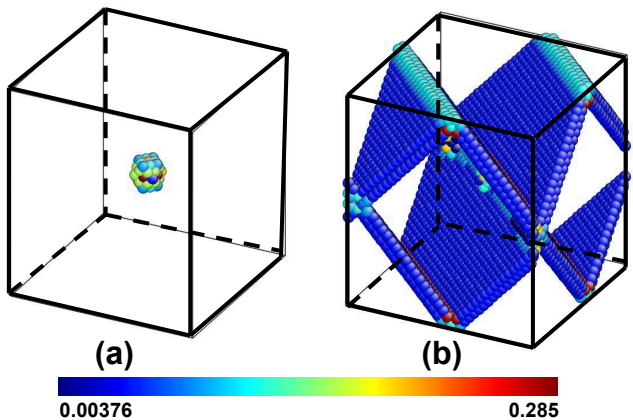}
\caption{CSP plot for the system immediately before (a) and after
(b) the yield point. The color bar represents the magnitude of CSP
values in the system. } \label{fig:v6}
\end{figure}

\begin{figure}[htp]
\includegraphics[width=\columnwidth,angle=0]{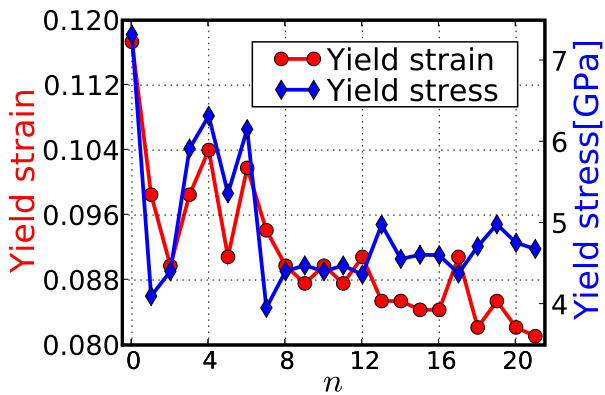}
\caption{The yield strain (circle, scaled to the left) and yield
stress (diamond, scaled to the right) of $n$-vacancy clusters under
uniaxial strain.}\label{fig:critical1D}
\end{figure}

\begin{figure}[htp]
\includegraphics[width=\columnwidth,angle=0]{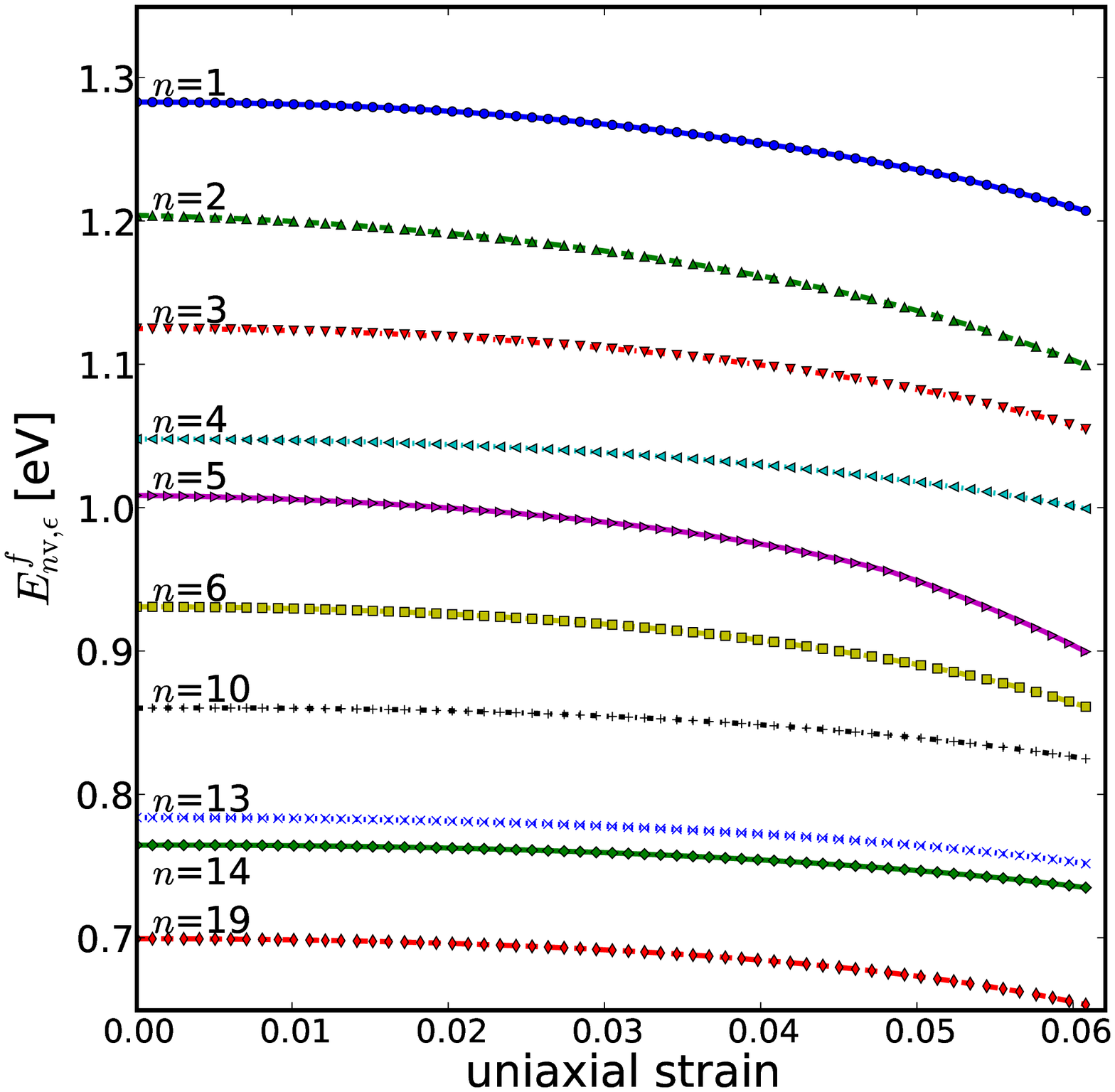}
\caption{The formation energy per vacancy
$E^{\rm{f}}_{n\rm{v},\epsilon}$ under uniaxial stain $\epsilon$ for
several representative vacancy clusters.}\label{fig:ca}
\end{figure}

We have calculated the formation energy
$E^{\rm{f}}_{n\rm{v},\epsilon}$ of the vacancy clusters under
uniaxial strain up to 0.062 beyond which some of the defect systems
start yielding. The results are summarized in \fig{fig:ca} for
representative vacancy clusters - the results of all vacancy
clusters have the same general features. It is found that the
formation energy decreases monotonically as a function of strain for
all clusters. This is because vacancy formation energy results from
the energy cost of bond breaking around the vacancies. By applying
tensile stress, the bonds are weakened owning to the stretching,
therefore it would cost less energy to break them comparing to the
strain-free case. The present result is consistent with that in Al
obtained by orbital-free density functional theory calculations
\cite{Gavini2008}. It appears that some of the energy curves tend to
converge (or cross) at larger strains, for example, $n=5, 6, 10$.
The apparent convergence suggests that a large stress can change the
relative stability of vacancy clusters of different sizes, which may
trigger structural changes among them.
\begin{figure}[htp]
\includegraphics[width=\columnwidth,angle=0]{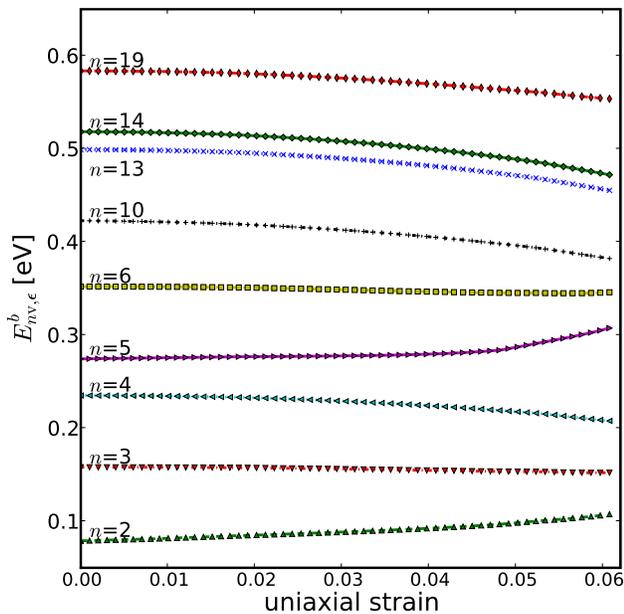}
\caption{The binding energy per vacancy
$E^{\rm{b}}_{n\rm{v},\epsilon}$ under uniaxial strain $\epsilon$ for
several representative vacancy clusters}\label{fig:cb}
\end{figure}

The binding energy per vacancy $E^{\rm{b}}_{n\rm{v},\epsilon}$ is
shown in \fig{fig:cb}. The binding energy remains positive for all
strains less than 0.062. For most of the vacancy clusters, the
binding energy decreases as a function of strain because the
attractive vacancies are pulled apart by the stress. It appears that
large strains could lead to break-up or re-arrangement of the
vacancy clusters.

\subsubsection{Isotropic volumetric tensile strain}
Next we study the behavior of the vacancy clusters under isotropic
volumetric tensile deformations. In this case of triaxial loading,
the volume of the system is no longer conserved, but with an equal
strain applied along $x$,$y$,$z$ directions simultaneously. At each
deformation step, the size of the simulation box is 1.0005 times
that of the previous step.
\begin{figure}[htp]
\includegraphics[width=\columnwidth,angle=0]{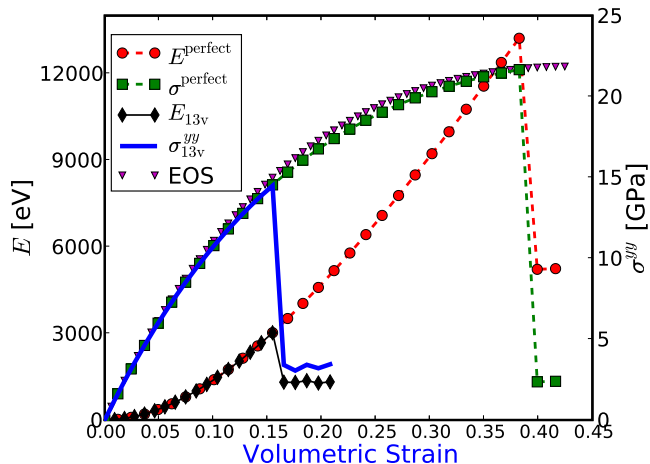}
\caption{The strain energy (scale to the left axis) and stress
(scale to the right axis) vs. isotropic volumetric strain for a
$13$-vacancy cluster (diamond and solid line) and perfect
lattice(circle and square) respectively. The triangle curve
represents the EOS from Eq. (13).} \label{fig:v6xyz}
\end{figure}

The strain energy and stress as a function of strain are shown in
Fig. 9, with $n=13$ as a representative because all vacancy clusters
behave in the same way. The energy increases with the strain before
the drop at a volume strain of $\epsilon=0.13$. The drop of the
energy is an indication of the onset of plasticity of the system.
The stress drops at the same strain as the energy. For the isotropic
volumetric tensile strain, $\sigma^{yy}=\sigma^{xx}=\sigma^{xx}=-P$,
and $P$ is the pressure. The non-linear behavior of the
stress-strain relation in Fig. 9 can be understood from the equation
of state (EOS) of solids \cite{Vinet89}:
\begin{equation}
\label{eq:eos1} P(V)= \frac{3 K_{0}(1-x)} {x^{2}} \exp [\xi (1-x)],
\end{equation}
where V$_0$ and $K_0$ are the volume and the bulk modulus of the
solid at equilibrium; $x\equiv(V/V_0)^{1/3}$,
$\xi\equiv\frac{3}{2}(K_0^{\prime}-1)$, and
$K_0^{\prime}\equiv\frac{\partial K_0}{\partial P}\vert_{P=0}$. For
the volumetric loading, $V=V_0 (1+\epsilon)$ and $x-1 \approx
\frac{1}{3}\epsilon$, thus we can rewrite \eqn{eq:eos1} as
\begin{equation}
\label{eq:eos2} \sigma(\epsilon)= \frac{K_0 \epsilon}
{(1+\epsilon/3)^2}  \exp [- \frac{\xi \epsilon}{3}].
\end{equation}
\eqn{eq:eos2} is plotted as the triangle curve in \fig{fig:v6xyz}
and it agrees very well with the EAM data by taking the experimental
value of $K_0=140$ GPa and $\xi=5.2$.

\begin{figure}[htp]
\includegraphics[width=\columnwidth,angle=0]{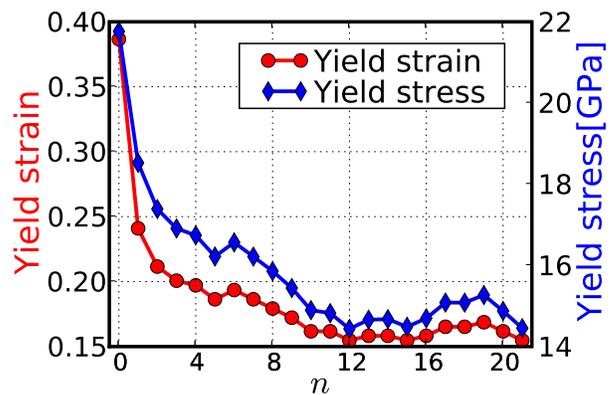}
\caption{The yield strain (circle, scaled to the left) and stress
(cross, scaled to the right) of $n$-vacancy clusters under the
volumetric tensile strain.}\label{fig:critical3D}
\end{figure}

The yield strain and yield stress for all vacancy clusters under the
volumetric deformation are plotted in \fig{fig:critical3D}. The
energy/stress curves of the defect systems trace closely to that of
the perfect lattice, indicating the vacancy clusters do not change
the behavior of the material before yielding. Overall, both the
yield stress and strain decrease as a function of $n$, suggesting
that vacancy clusters can drastically reduce the yield strength of
the material. It should be noted that the quantitative values of the
yield stress/strain obtained here do not correspond to the realistic
situation due to the high vacancy concentrations in the simulations;
however, the qualitative features remain to be valid.

\begin{figure}[htp]
\includegraphics[width=\columnwidth,angle=0]{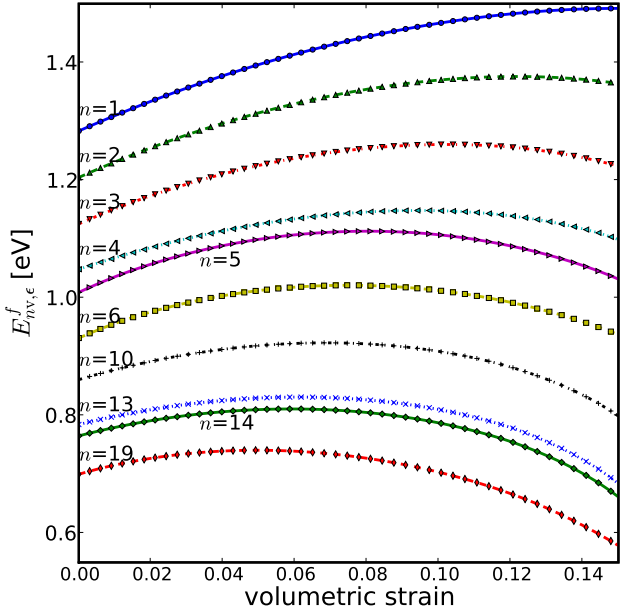}
\caption{The formation energy per vacancy
$E^{\rm{f}}_{n\rm{v},\epsilon}$ under the volumetric strain
$\epsilon$ for several representative vacancy
clusters.}\label{fig:c3a}
\end{figure}

The formation energy per vacancy under the volumetric tensile strain
is shown in \fig{fig:c3a}. Overall, we find that the formation
energy first increases with the strain to a maximum and then
decreases; this is different from the uniaxial case. The difference
is due to the different loading conditions: in the uniaxial case,
the volume of the system is fixed, therefore to a good
approximation, the surface area of the void is also fixed. On the
other hand, the surface area of the void increases as the volume
increases in the triaxial case, therefore the inner surface energy
increases as a function of the tensile strain. The increasing
surface energy dominates the decreasing bonding energy contribution
at small strains, hence the formation energy reaches a maximum at an
immediate strain. For small vacancy clusters, such as $n=1$ and
$n=2$, the surface energy contribution is negligible owing to the
small size of the vacancy clusters, thus their formation energy
increases monotonically.

For the volumetric strain, the binding energy per vacancy increases
monotonically with the strain, and it increases faster when $n$ is
larger. Such a behavior can be understood again from the liquid-drop
model. Under the volumetric strain $\epsilon$, the volume of a
monovacancy is $v_0(1+\epsilon_{11})^3$ where the linear strain
$\epsilon_{11} \approx \epsilon/3$. Thus the two coefficients in Eq.
(11) become $c_1(1+\epsilon_{11})^2$ and $c_2(1+\epsilon_{11})$
respectively. $E^{\rm{b}}_{n\rm{v},\epsilon}$ can then be expressed
as a quadratic function of $\epsilon_{11}$ or $\epsilon$:
\begin{eqnarray}
\label{eq:liqidb3}
 E^{\rm{b}}_{n\rm{v},\epsilon} &=&
c_1(1-n^{-\frac{1}{3}})(1+\epsilon_{11})^2 - c_2
(1-n^{-\frac{2}{3}})(1+\epsilon_{11})  \nonumber \\
&=&c_1(1-n^{-\frac{1}{3}})\epsilon_{11}^2  \nonumber \\
&+&[2c_1(1-n^{-\frac{1}{3}})-c_2(1-n^{-\frac{2}{3}})] \epsilon_{11} \nonumber \\
&+&c_1(1-n^{-\frac{1}{3}})-c_2(1-n^{-\frac{2}{3}}).
\end{eqnarray}
Since $c_1(1-n^{-\frac{1}{3}})>0$ and
$[2c_1(1-n^{-\frac{1}{3}})-c_2(1-n^{-\frac{2}{3}})]>0$,
$E^{\rm{b}}_{n\rm{v},\epsilon}$ would increase monotonically with
the volumetric strain $\epsilon$. The larger the $n$ value, the
larger the coefficient of the quadratic term
$c_1(1-n^{-\frac{1}{3}})$, hence $E^{\rm{b}}_{n\rm{v},\epsilon}$
increases faster with respect to the strain.

\begin{figure}[htp]
\includegraphics[width=\columnwidth,angle=0]{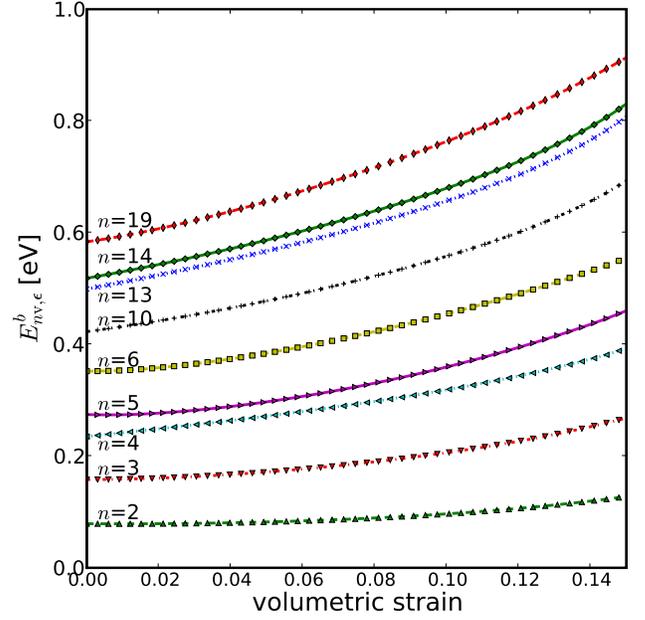}
\caption{The binding energy per vacancy (eV)
$E^{\rm{b}}_{n\rm{v},\epsilon}$ under volumetric strain $\epsilon$
for several representative vacancy clusters.}\label{fig:c3b}
\end{figure}

\subsection{Thermodynamic Stability}
Finally, we address the thermodynamic stability of the vacancy
clusters. At finite temperatures $T$, a vacancy cluster could
dissociate into smaller clusters, or eventually into monovacancies.
In the following, we use the concept of total probability of
dissociation $P_{\rm tot}(n)$ \cite{Zhang2008} to analyze the
thermal stability of the vacancy clusters. $P_{\rm tot}(n)$ can be
expressed as $P_{\rm tot} = \sum_{i} P_{i}(n)$ where $P_{i}(n)$
represents the probability of a particular (the $i$th) way of
dissociation. For example, an $n$-vacancy cluster may dissociate
into $n$ monovacancies simultaneously, whose probability, $P_{1}(n)$
is given by
\begin{eqnarray}
\begin{split}
P_{1}(n)&=\frac{(c_{\rm
1v})^{n}}{c_{n\rm{v}}}=\frac{n}{g_{n\rm{v}}}e^{-({n}G_{\rm
1v,\epsilon}^{\rm
F}-G_{n\rm{v},\epsilon}^{\rm F})/k_{\rm B}T}\\
&=\frac{n}{g_{n\rm{v}}}e^{-G_{n\rm{v},\epsilon}^{\rm B}/k_{\rm B}T};
\end{split}
\end{eqnarray}
or an $n$-vacancy cluster could break into a monovacancy plus an
$(n-1)$-vacancy cluster, whose probability $P_{2}(n)$ is expressed
as
\begin{eqnarray}
\begin{split}
P_{2}(n)=\frac{c_{\rm 1v}c_{(n-1)\rm{v}}}{c_{n\rm{v}}}=\frac{ng_{
(n-1)\rm{v}}}{(n-1)g_{n\rm{v}}}e^{-(G_{n\rm{v},\epsilon}^{\rm
B}-G_{(n-1)\rm{v},\epsilon}^{\rm B})/k_{\rm B}T}.
\end{split}
\end{eqnarray}
where $g_{n\rm{v}}$ is the coordination number of the $n$-vacancy
cluster \cite{Grimvall1999}, etc. The concentration of the
$n$-vacancy cluster is denoted by $c_{nv}$ and $k_{\rm B}$ is the
Boltzmann constant. The definition of $P_{i}(n)$ can be used in both
thermal equilibrium condition and a constant supersaturation
condition \cite{Zhang2008}. Here we have to consider all possible
dissociation paths in calculating a given $P_{\rm tot}(n)$. The
variation of $G_{n\rm{v},\epsilon}^{\rm B} $ with respect to $n$
determines the dissociation probability \cite{Zhang2008}; while an
$n$-vacancy cluster is thermally stable under a positive and rapidly
increasing $G_{n\rm{v},\epsilon}^{\rm B}$ function, a negative and
slowly increasing $G_{n\rm{v},\epsilon}^{\rm B}$ would result in a
high dissociation probability of the $n$-vacancy cluster.

\begin{figure}[htp]
\includegraphics[width=\columnwidth,angle=0]{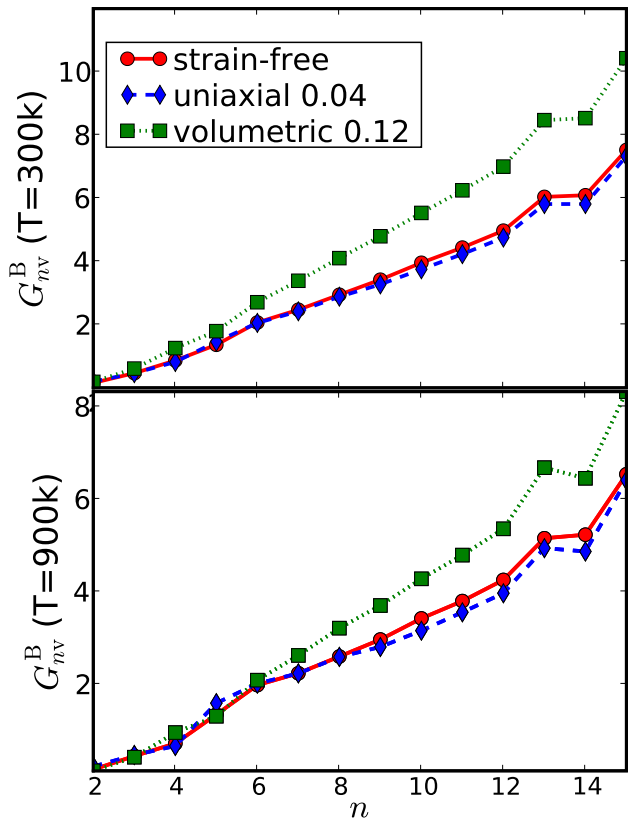}
\caption{The Gibbs binding free energy of $n$-vacancy clusters at
different temperatures for strain-free (circle), uniaxial tensile
strain (diamond) and volumetric tensile strain
(square).}\label{fig:Gnv}
\end{figure}

With Eqs.(5)-(7), we can evaluate the binding entropy
$S_{n\rm{v},\epsilon}^{\rm B}$, and then the Gibbs binding free
energy of an $n$-vacancy cluster $G_{n\rm{v},\epsilon}^{\rm B}$ at a
given temperature $T$. In \fig{fig:Gnv}, we present
$G_{n\rm{v},\epsilon}^{\rm B}$ at $T=300{\rm K}$ and $T=900{\rm K}$
under the uniaxial and volumetric tensile strains. From the Gibbs
binding free energy, one can estimate the total dissociation
probability $P_{\rm tot}(n)$ of an $n$-vacancy cluster. The
calculation reveals that the total dissociation probabilities of all
$n$-vacancy clusters, except $n=14$, are much less than 0.5 under
both strain-free and volumetric tensile strain conditions, which
means they are thermodynamically stable. On the other hand, under
the volumetric strain, the 14-vacancy cluster has more than $50\%$
probability to dissociate into a 13-vacancy cluster and a
monovacancy because $G_{14\rm{v},\epsilon}^{\rm B}$ is smaller than
$G_{13\rm{v},\epsilon}^{\rm B}$. Under a uniaxial tensile strain,
10- and 11-vacancy clusters are not stable at $T=900{\rm K}$; the
10- or 11-vacancy cluster has a high probability to dissociate into
two 5-vacancy clusters or one 5- and one 6-vacancy clusters,
respectively.

\section{CONCLUSION}
We have examined structure, mechanical and thermodynamic stability
of $n$-vacancy clusters where $n$ goes from 2 to 21. We determine
the most stable atomic structure of the vacancy clusters based on
simple intuitive considerations and energetic calculations. More
extensive (such as global search) and accurate (such as quantum
mechanical) calculations are required to settle the issue with more
confidence. The present work should only be considered as a
preliminary effort. We find that the formation energy per vacancy
decreases while the binding energy increases as a function of
cluster size $n$. We probe the mechanical stability of the vacancy
clusters by applying uniaxial and volumetric tensile strains. It is
observed that yield stress and yield strain of the material are
significantly reduced by the vacancy clusters. However, the presence
of the vacancy cluster does not change the elastic behavior of the
material before yielding. We find that formation energy per vacancy
decreases as a function of uniaxial strain for all clusters, while
the binding energy shows a more complicated behavior. For volumetric
deformations, the formation energy per vacancy increases first then
decreases as a function of strain. The increase of the formation
energy at smaller strains is due to the increased surface energy
associated with the void. We determine the thermodynamic stability
of the clusters by calculating the Gibbs free binding energy and
resultant probability of dissociation. We find that most of the
vacancy clusters under study are thermodynamically stable except the
14-vacancy cluster, which has a high probability of dissociating
into a 13-vacancy cluster and a monovacancy. In addition, at 900K
the 10-and 11-vacancy clusters will dissociate into two 5-vacancy
clusters or a 5- and a 6-vacancy clusters, respectively under a
uniaxial strain.

\begin{acknowledgments}
The work at California State University Northridge was supported by
NSF PREM grant DMR-0611562 and the ACS Petroleum Research Funds
PRF43993-AC10.
\end{acknowledgments}


\end{document}